# Nonreciprocal Emergence of Hybridized Magnons in Magnetic Thin Films

*Wenjie Song[#], Xiansi Wang[#], Chenglong Jia[\*], Xiangrong Wang, Changjun Jiang, Desheng Xue, and Guozhi Chai[\*]*

Wenjie Song, Prof. Chenglong Jia, Prof. Changjun Jiang, Prof. Desheng Xue, and Prof. Guozhi Chai

Key Laboratory for Magnetism and Magnetic Materials of the Ministry of Education, Lanzhou University, Lanzhou 730000, People's Republic of China

Dr. Xiansi Wang

School of Physics and Electronics, Hunan University, Changsha 410082, China

Prof. Xiangrong Wang

Department of Physics, The Hong Kong University of Science and Technology, Clear Water Bay, Kowloon, Hong Kong, People's Republic of China

HKUST Shenzhen Research Institute, Shenzhen 518057, People's Republic of China

\# These authors contributed equally to this work

\* E-mail: cljia@lzu.edu.cn, chaigzh@lzu.edu.cn

**Magnons (the quanta of spin waves), unlike electrons, do not suffer from Ohmic losses which makes them an attractive information carrier in communication and data processing. The possibility of non-reciprocal magnon propagation is of great practical relevance to directional communications, particularly when the nonreciprocity is steerable through magnons themselves in a monologic magnonic device. This work demonstrates explicitly such nonreciprocal magnons in $Ni_{79}Fe_{21}$ thin films. Evidences of nonreciprocal emergence of hybridized dipole-exchange spin waves at two surfaces of $Ni_{79}Fe_{21}$ nano-films deposited on surface oxide silicon substrate are provided by studying magnon transmission and asymmetry, via Brillouin light scattering measurements. The dipole-dominated spin wave and exchange-dominated spin wave are found to be localized near the top and bottom surfaces, respectively, and traveling along opposite directions. The nonreciprocity and the localization can be tuned by an in-plane magnetic field. Our findings provide a simple and flexible approach to nonreciprocal all-magnon logic devices with highly compatible with silicon-based integrated circuit technology.**

Magnons, or spin waves, are low-energy excitations of magnetic materials that can be generated by various ways. Their dispersion and propagation characteristics can be tailored through material engineering down to the nanoscale thin films.[1–4] Tailors properties render possible realization of magnonic circuits based on wave interferences and interactions. Various applications have been demonstrated including logic gates,[5,6] filters,[7] waveguides,[8,9] diodes [10], beam-splitters[11, 12] and multiplexors [13]. Generally speaking, as demonstrated in Fig. 1, two types of spin waves exist: short-wavelength exchange spin waves and relatively long-wavelength magnetostatic spin waves. In contrast to the symmetric exchange spin waves dominated by strong exchange interactions, the magnetostatic surface spin waves (MSSWs) or the Damon-Eshbach (DE) spin waves that are governed by the long-range dipole-dipole interactions are particularly interesting because of their nonreciprocal property, [14–16] i.e., they propagate only in the direction of the outer product of the magnetization direction and surface normal and, therefore, localize themselves on the opposite film surfaces. Such nonreciprocal MSSWs with lack of the time-reversal and space-inversion symmetries have been well studied[17, 18], however there is no clear experimental observation of how the nonreciprocity of MSSW behaves when it is hybridized with other reciprocal spin wave modes, an issue of key importance for the development of on-single-chip, all-magnon devices.

Magnetic films are useful for investigating the interplay of nonreciprocal MSSWs with spin standing waves (SSWs) determined by the exchange interaction and film thickness.[19, 20] Noticing the penetration depth of MSSWs at the scale of their wavelength, the nonuniform spatial profile across the thickness is negligible for very thin films (several nanometers), and only one spin-wave mode is observed at GHz frequencies. For thicker films, the vertical confinement gives rise to longer wavelengths and lower eigen-frequencies of PSSWs. Consequently, PSSWs starts to become relevant for the magnon dynamics. When the spectrums of PSSW and MSSW modes intersect, they strongly couple with each other and form hybridized dipole-exchange modes.[21, 22] There are several experiments with the aim to characterize the dipolar-exchange modes.[23–25] It has been shown that when the frequencies of MSSW modes are far from the PSSWs, the MSSWs can be very robust against defects.[26] In the opposite limit when the frequencies of MSSWs and PSSWs are very close, multiple dipole-exchange bands were observed, and an asymmetry in the intensities of Stokes and anti-Stokes branches can be identified, although the converse is not very obvious.[24] Some highly nonreciprocal spin waves and magnon-magnon coupling were observed in magnetic metal/insulator heterostructures,[27–31] layered crystal, [32] compensated ferrimagnet,[33] synthetic antiferromagnets,[34] and ferromagnetic films with interfacial Dzyaloshinskii-Moriya interaction.[35, 36] However, an unambiguous evidence of the nonreciprocal hybridization between MSSW and PSSW modes is still missed in experiments and a direct all-magnonic control of hybridized spin waves in single magnetic layer is yet to be clearly demonstrated. In particular, as for advanced applications, a material choice is important. A device based on widely available magnetically-soft materials such as $Ni_{79}Fe_{21}$ (permalloy) and compatible with silicon-based integrated circuits would be highly advantageous.

Our aim here is to realize nonreciprocity transfer from the MSSW to PSSW modes in in-plane magnetized $Ni_{79}Fe_{21}$ films. Note that in the direction of experimental clarification, the thickness of films is prepared to have only the first PSSW mode excited within the frequency and wave vector range accessible to the measurement (the higher order PSSWs are lifted to much higher frequencies). As shown in **Figure 1**, the magnon-magnon coupling between MSSW and PSSW allows for space-inversion symmetry breaking of hybridizations, leading to nonreciprocity and unidirectional invisibility of two dipole-exchange eigenmodes. This nonreciprocal emergence of hybridized magnons is unambiguously observed in our experimental setup by adjusting the thickness of $Ni_{79}Fe_{21}$ films and agrees well with the dispersion characteristics derives from a two-band model with a coupling strength ~ 1 GHz. The detailed micromagnetic simulations confirm the symmetry breaking and unravel further that the coupled MSSW and PSSW branches are indeed localized on the two opposite surfaces and propagate in the opposite directions. [37] The localization and travelling direction of spin waves are intertwined, and changes their directions simultaneously by rotating the in-plane magnetic field. In this way we demonstrate a flexible scheme of all-magnonical control of nonreciprocity.

We deposited FeNi films on surface oxide 400 um thick Si substrates by radio frequency magnetron sputtering at room temperature. The oxide surface can enhance the signal of scattering light in Brillouin Light Scattering (BLS) measurements.[38] The thicknesses of FeNi layer are controlled by varying the sputtering time. The static magnetic properties of FeNi films were measured by a vibrating sample magnetometer (VSM). The measured in-plane magnetic hysteresis loops of FeNi (34 nm) and FeNi (44 nm) films indicate that the in-plane anisotropy fields are less than 12 Oe (Figure S1, Supporting Information), which is much smaller than an external in-plane magnetic field **H** = 500 Oe. In the following, the films are considered as isotropic in the film (*xy*) plane.

In experiment, the spin wave spectra in $Ni_{79}Fe_{21}$ films are measured by using wave-vector- and frequency-resolved BLS technique at room temperature (Figure 1a). The static magnetization $\mathbf{m}_0$ is aligned along the +*y* axis by the applied magnetic field **H** = 500 Oe. The BLS incident laser and scattering light are performed in the 180°-backscattering geometry, and based on the (3+3)-pass tandem Fabry-Pérot interferometer, which is effective for achieving the frequency resolution of surface spin waves.[39] With backscattering geometry, a laser beam with the wavelength λ = 532 nm is focused onto the $Ni_{79}Fe_{21}$ films, and the photons are inelastically backscattered by the magnons in the vicinity of the top surface within the laser penetration depth: the photon loses its kinetic energy to create one magnon (Stokes process) or gains energy by absorbing one magnon (anti-Stokes process). Based on the laws of conservation of momentum and energy, the propagating direction and energy dispersive spectra of magnons can be uniquely determined by the BLS measurements. The incident plane of the laser light is perpendicular to the *y* axis. Thus, the measured in-plane wave vector $k_\parallel$ is along the x direction. In the scattering process, the Stokes (anti-Stokes) peaks in BLS spectra correspond to $k_\parallel = 4\pi \sin\theta/\lambda$ along the +*x* (−*x*) direction, where the *θ* is the laser-light incident angle. By varying *θ*, the magnon frequency of different wavevectors can be obtained. The measured intensities versus frequency for different in-plane wavevectors of FeNi (34 nm) and FeNi (44 nm) films are shown in **Figure 2a and 3a**, respectively, where the frequency resolution is 0.07 GHz and the range of measurable wavevectors is between ± 17 rad/*μ*m.

The spin dynamics in our setting are well described by the phenomenological Landau-Lifshitz-Gilbert (LLG) equation,[40]

$$\frac{\partial \mathbf{m}}{\partial t} = -\gamma \mathbf{m} \times \left(\frac{2A}{M_s}\nabla^2 \mathbf{m} + \mathbf{H} + \mathbf{h}\right) + \alpha \mathbf{m} \times \frac{\partial \mathbf{m}}{\partial t}, \quad (1)$$

Where $\mathbf{m}$ is the magnetization vector, $\gamma$ is the gyromagnetic ratio, $A$ is the exchange constant, and $\mathbf{h}$ is the dipolar field. In the magnetostatic limit and low damping limit, by ignoring the exchange interaction (or the nonlocal dipolar interaction), the well-known MSSW (or PSSW spectra) can be obtained from the linearized LLG equation.[19, 20] The frequency of MSSW $f_M$ and n-th PSSW $f_P^n$ are

$$f_M = \gamma \sqrt{H(H + 4\pi M_s) + 4\pi^2 M_s^2 \left(1 - e^{-2|k_\parallel| t_M}\right)}, \quad (2)$$

$$f_P^n = \gamma \sqrt{(H + H_{ex,n})(H + H_{ex,n} + 4\pi M_s)}, \quad (3)$$

where $H_{ex,n} = \frac{2A}{M_s}\left[\left(\frac{n\pi}{t_M}\right)^2 + k_\parallel^2\right]$, $M_s$ is the saturation magnetization, $t_M$ is the film thickness, and $k_\parallel$ is the transferred in-plane spin wave vector.

We first look into the BLS results of a 34-nm-thick Ni$_{79}$Fe$_{21}$ film in the presence of an applied magnetic field $\mathbf{H}$ = 500 Oe. For comparison, the density plot of BLS intensity in ω-$k$ plane are separately shown in Figure 2b for Stokes processes with magnons (at the bottom surface) propagating away from the incoming laser beam and Figure 2c for anti-Stokes processes with magnons (at the top surface) propagating in the opposite direction. Using parameters $\gamma$ = 2.82 MHz/Oe, $4\pi M_s$ = 1.04×10$^4$ G (confirmed by separate hysteresis loops with VSM), $A$ = 1.1 ×10$^{-6}$ erg/cm, and $t_M$ = 34 nm, the measured spectra agree very well with the dispersion relationship Equation 2 and 3 for MSSWs and PSSWs with $n = 1$, as shown in Figure 2d and e. It is clear that both nonreciprocal MSSWs and symmetrical PSSWs are excited but no hybridization occurs in the range of the wave vector ($|k_\parallel| \leq 17$ rad/$\mu$m), implying a relatively weak magnon-magnon coupling ($g < 3$ GHz) in $Ni_{79}Fe_{21}$ films (also confirmed with the simulating results shown in Figure S2, Supporting Information).

To reach the vicinity of the crossing point of MSSW and 1st PSSW bands, $Ni_{79}Fe_{21}$ film of 44-nm-thick is used. The measured intensities versus frequency $f$ for different $k_\parallel$ are shown in Figure 3a. The density plots in $k - f$ plane for Stokes and anti-Stokes branches are presented in Figure 3b and c, respectively. Clearly, the MSSW band and the 1st PSSW band of the 44-nm $Ni_{79}Fe_{21}$ film intersect with each other. One obtains the hybridized dipole-exchange modes from an effective coupled two-band Hamiltonian, $\mathcal{H} = \begin{pmatrix} f_M & g/2 \\ g/2 & f_P^1 \end{pmatrix}$. The two hybridized dipole-exchange eigenmodes are

$$f_\pm = \frac{1}{2}(f_M + f_P^1) \pm \frac{1}{2}\sqrt{(f_M - f_P^1)^2 + g^2}, \tag{4}$$

which agree with the experimental peak positions (symbols), as demonstrated in Figure 3d and e with the coupling strength $g = 1$ GHz (fitted curves). The BLS results of 40-nm-thick $Ni_{79}Fe_{21}$ film are presented as a comparison. (Figure S3, Supporting Information). The calculated hybridization strength of 40-nm $Ni_{79}Fe_{21}$ film is approximately $g = 1.6$ GHz, which is stronger than the hybridization strength of 44-nm $Ni_{79}Fe_{21}$ film.

Furthermore, upon hybridization the resultant dipole-exchange spin wave modes show a strong nonreciprocal behavior. Figure 3a-c evidence that before the crossing point ($|k_\parallel| < 10$

rad/$\mu$m), both MSSW and PSSW modes are excited and PSSW itself possesses symmetric exchange dispersion of $k_\parallel$. However, after the anti-crossing, only MSSW branch (PSSW branch) can be detected in the anti-Stokes (Stokes) side. Note that the anti-Stokes (Stokes) peaks correspond to wave-vector along the $+x$ ($-x$) direction in our geometry. With $\hat{\mathbf{n}}$ being the normal of the top surface (the surface facing the incident laser), the spin wave nonreciprocity can be summarized as follows: the MSSW branch propagates as expected along $\mathbf{m}_0 \times \hat{\mathbf{n}}$, while the PSSW branch becomes nonreciprocal as well and travels only along the opposite direction at the opposite surface. The above unidirectional propagating property was double checked by rotating the in-plane field $\mathbf{H}$ along $-y$ so that $\mathbf{m}_0$ reversed its direction. Accordingly, the patterns of the Stokes and anti-Stokes intensities are interchanged (Figure S4, Supporting Information).

In order to obtain a full picture of the observed nonreciprocity, we solve the spin wave spectra numerically in a more precise way. Following the approach in Refs. 21 and 41, we adopt the magnetostatic assumption (i.e. ignore the electric part of the Maxwell's equations), so that the dipolar field can be written as the gradient of a scalar potential $\varphi$, Then we expand $\mathbf{m}$ around $\mathbf{m}_0$ as $\mathbf{m} = \mathbf{m}_0 + \delta\mathbf{m}$ and keep only linear terms of $\delta\mathbf{m}$ in the LLG equation (Equation 1). The divergenceless property of the magnetic flux density $\mathbf{B}$ can be written as $\nabla \cdot \mathbf{B} = \nabla \cdot (\mathbf{h} + \mathbf{H} + 4\pi M_s \mathbf{m}) = 0$, giving rise to $\nabla \cdot (\mathbf{h} + 4\pi M_s \delta\mathbf{m}) = 0$. Together with the boundary conditions, the spin wave dispersion relations are obtained (gradient blue and red dashed lines) in **Figure 4a**, which are the exact dipole-exchange modes we interested in. To explain the nonreciprocal behavior, we further solve the wavefunction amplitudes of $\mathbf{m}$ components across the thickness direction. We consider $k_\parallel$ = -12 rad/μm, indicated by two orange circles in Figure 4a. The amplitudes of $\mathbf{m}_x$ and $\mathbf{m}_z$ of the MSSW branch (the higher-frequency band) are demonstrated in Figure 4b. The dynamic $\delta\mathbf{m}$ components are mainly localized at the bottom surface, which is consistent with the "pure" MSSW according to the Damon-Eshbach theory.[14] The amplitudes of $\mathbf{m}_x$ and $\mathbf{m}_z$ of the PSSW branch (the lower-frequency band) are given in Figure 4c. Unlike the "pure" PSSW modes which are inversion-symmetric, the hybridized mode breaks the space-inversion symmetry and is localized at the top surface. For positive $k_\parallel$ = 12 rad/μm, the wavefunction amplitudes are just the mirror images of Figure 4b and c because the film is $\mathbb{C}_2$ symmetric around *y* axis. As laser stimulating and detecting the spin waves impinges from the top surface, only the modes with sufficiently large amplitudes at the top surface scatter from light and are detected (Figure 1c-e). This is why in the experiment only positive-*k* MSSW branch and negative-*k* PSSW branch are observed.

To further substantiate our explanation, we performed micromagnetic simulations using the Mumax[3] package with the material parameters mentioned above.[42] The 40960 × 3200 × 44 nm[3] large system is considered with mesh size of 10 × 100 × 1.375 nm[3]. As we are only interested in the wave propagating in *x* direction and the wavefunction distribution in *z* direction, we can save on the computation time and perform 32 repeats of a single mesh in *y* direction. In the calculations the spin waves are excited by a finite temperature *T* = 10 K. The Fourier transform of $\mathbf{m}_z$ at the top surface is plotted in Figure 4a as a comparison. We find that the experimental data, theoretical bands and micromagnetic simulation results show an excellent agreement: not only the spin wave spectra, but also the strong nonreciprocal behavior is consistently reproduced.

Taking advantage of the micromagnetic simulations with precisely adjustable parameters, the effects of two involved magnetic interactions, i.e., exchange interaction and dipole-dipole interaction are investigated in detail (Figure S5, Supporting Information). It has been found that the mode hybridization is affected by the strength of dipole-dipole and exchange interactions, however, the nonreciprocal behavior is subtle and experimentally evident only in certain real materials. $Ni_{79}Fe_{21}$ films are proved to be the right choice. As a comparison, density plots of measured BLS spectra of $Co_{20}Fe_{60}B_{20}$ (40 nm) film comparing with the theoretical curves are shown in **Figure 5**. Even though $Co_{20}Fe_{60}B_{20}$ has larger saturation magnetization and thus stronger dipole-dipole interactions, the nonreciprocal hybridization phenomenon is vague and imprecise in $Co_{20}Fe_{60}B_{20}$ film. We also carried out further studies on Co (44 nm) film with the largest saturation magnetization and highest exchange interactions among three types of metallic ferromagnetic films. The hybridized dipole-exchange spin waves in Co film, however, presents a moderate non-reciprocity (Figure S6, Supporting Information).

In summary, by varying the thickness of $Ni_{79}Fe_{21}$ films, we demonstrated a controllable, yet simple and flexible, all magnon setup for realizing nonreciprocity transfer between spin waves. The hybridized dipole-exchange spin waves with a large gap of about 1 GHz are obtained. Not only the MSSW modes but also the exchange-dominated PSSW branch are found to be localized on the film surface, and possess strongly nonreciprocal behavior, which can be further tuned by an external applied in-plane magnetic field. These effects are well explained by the magnetostatic theory and can be quantitatively reproduced by micromagnetic simulations.

Obviously, our scheme is fundamentally different from cavity-based magnonics, in which the coherent and dissipative interaction between microwave cavity modes and spin wave modes is a key ingredient for the development of wave nonreciprocity. The system presented here is a simple way for achieving strongly nonreciprocal, and controllable spin waves with clear potential for integration in all-magnonic devices.

**Supporting Information**

Supporting Information is available from the Wiley Online Library or from the author.

**Acknowledgements**

The authors thank J. Berakdar, Wanjun Jiang, Haiming Yu, and Jilei Chen for their fruitful discussions. This work is supported by the National Natural Science Foundation of China (NSFC) (Nos. 51871117, 91963201 and 11834005), the Natural Science Foundation for Distinguished Young Scholars of GanSu Province, China (No. 20JR10RA649) and the Program for Changjiang Scholars and Innovative Research Team in University (No. IRT-16R35). X. S. W. acknowledges the support from the Natural Science Foundation of China (NSFC) (Grant No. 11804045). X.R.W acknowledge supports from Hong Kong RGC (Grants No.16301518, 16301619 and 16300117).

**Competing Interests statement**

The authors declare no competing interests.

Received: ((will be filled in by the editorial staff))
Revised: ((will be filled in by the editorial staff))
Published online: ((will be filled in by the editorial staff))

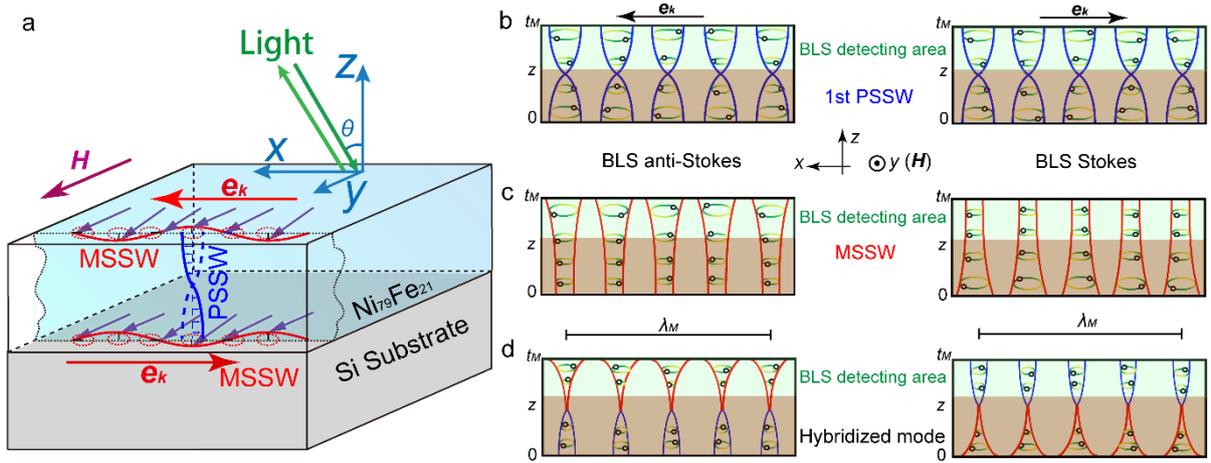

**Figure 1.** a) Schematics of hybridization of MSSW and 1st-PSSW modes of BLS measurement geometry. b), c) and d) Detailed cross-sectional profiles of 1st PSSWs (b), MSSWs (c) and hybridized spin waves (d) in the film. The green-yellow circles indicate the coherent spin precession and corresponding amplitude. As can be seen, all the spin waves break the time-reversal symmetry due to their right-handed procession. (b) The pure PSSWs are reciprocal: Both the wave dispersion and the propagation velocity are symmetric for opposite propagation directions. However, (c) MSSWs is inherently nonreciprocal because they lack the space-reversal symmetry along the z-direction as well. (d) The hybridized dipole-exchange spin waves break space-inversion symmetry and result in nonreciprocity transfer from the MSSW to PSSW modes. The upper light-green areas in b, c, and d demonstrate the detectable region of BLS measurements. $\lambda_M$ is the wavelength which can be determined by the spatial phase differences.

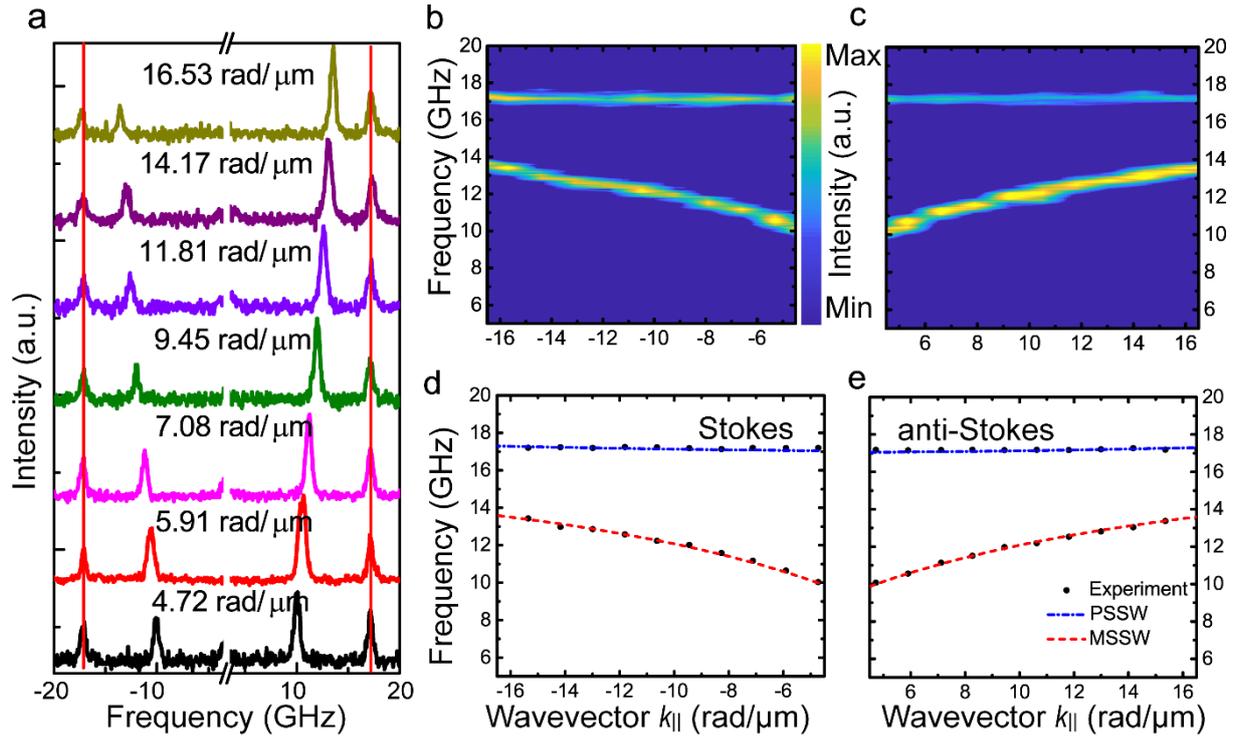

**Figure 2.** a) BLS intensities spectra of $Ni_{79}Fe_{21}$ (34 nm) film for different wavevectors $k_{\parallel}$ with **H** = 500 Oe. Red vertical lines corresponding to 1st PSSWs. Density plot of BLS intensity in ω-k plane: b) Stokes and c) anti-Stokes. Dependence of the spin waves frequency on the transferred wave vector $k_{\parallel}$: d) Stokes and e) anti-Stokes. The dashed red line is the MSSW band $f_M$. The dash-dotted blue line is the 1st PSSW $f_P^1$.

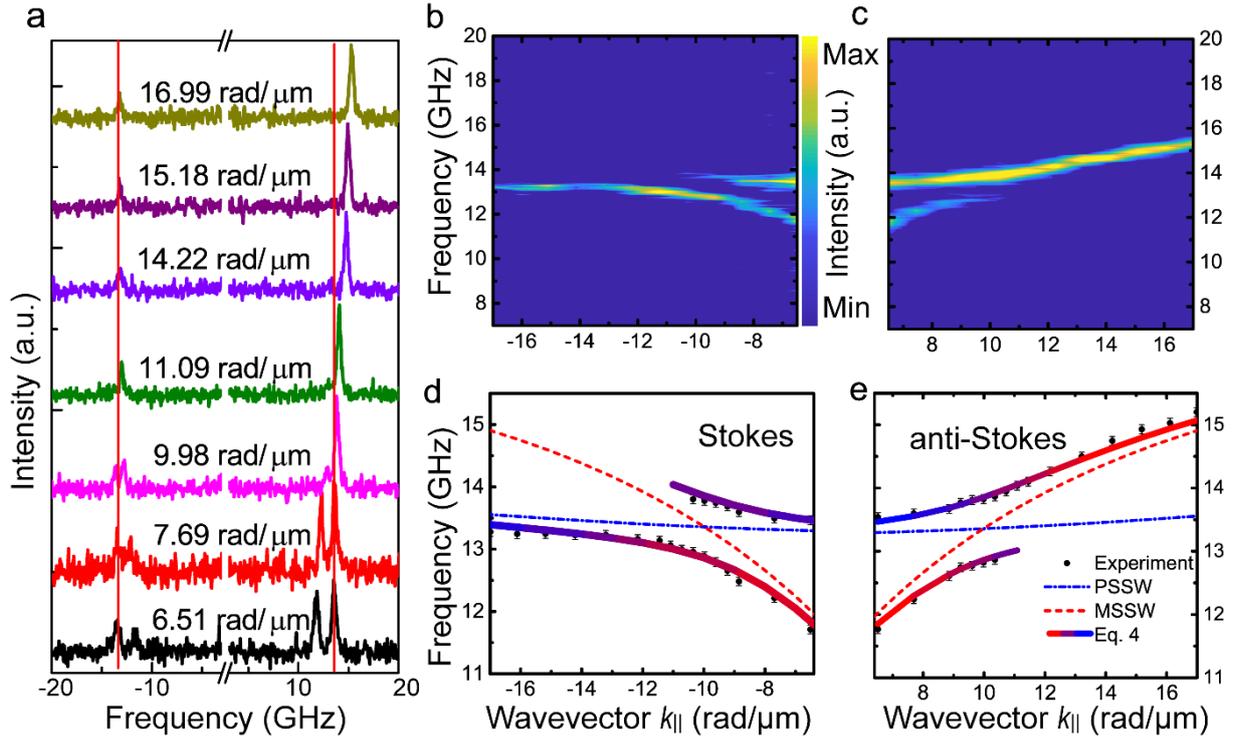

**Figure 3.** a) BLS intensities spectra of $Ni_{79}Fe_{21}$ (44 nm) film for different wavevectors $k_{\parallel}$ with **H** = 500 Oe. Red vertical lines corresponding to 1st PSSWs. Density plot of BLS intensity in ω-*k* plane: b) Stokes and c) anti-Stokes. Dependence of the spin waves frequency on the transferred wave vector $k_{\parallel}$: d) Stokes and e) anti-Stokes. The dashed red line is the MSSW band $f_M$. The dash-dotted blue line is the 1st PSSW $f_P^1$. The gradient blue and red lines are fitting curves using the two-band model.

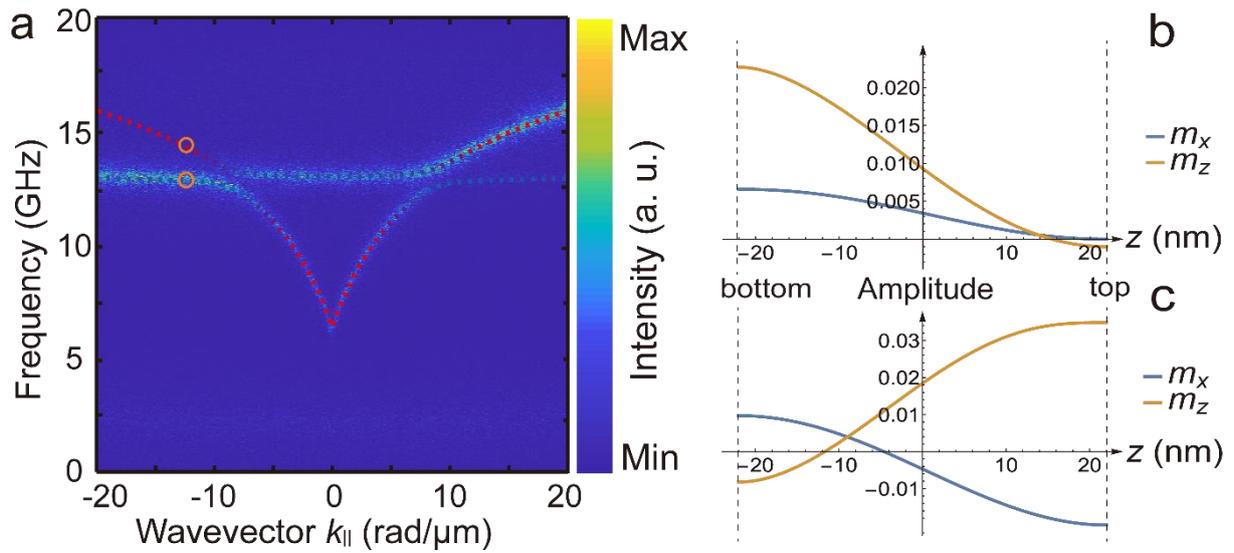

**Figure 4.** a) The simulation results of Ni$_{79}$Fe$_{21}$ (44 nm) film. The gradient blue and red lines are the theoretically derived dipole-exchange bands. The density plot is the micromagnetic simulation result. b) and c) The wavefunction amplitudes of $k_\parallel = -12$ rad/$\mu$m of (b) the higher-frequency band (MSSW branch) and (c) the lower-frequency band (PSSW branch).

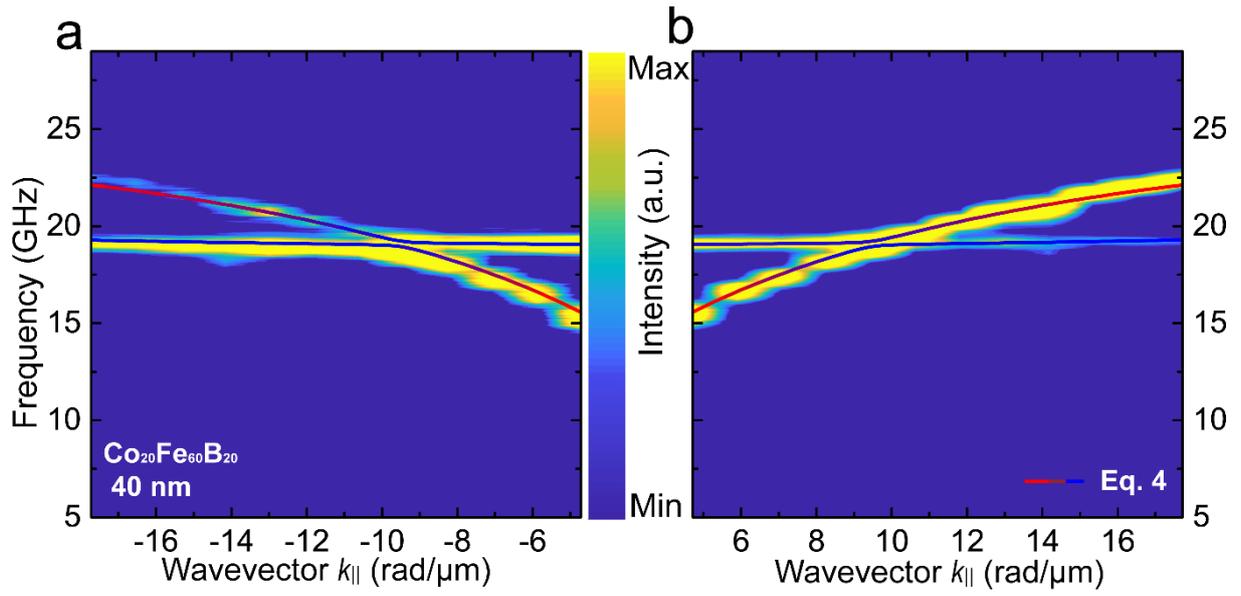

**Figure 5.** a) and b) Density polt of BLS intensities spectra of $Co_{20}Fe_{60}B_{20}$ (40 nm) film in $\omega$-$k$ plane with **H** = 500 Oe: a) Stokes and b) anti-Stokes. The gradient solid lines are given based on the two-band model (Equation 4).